\documentclass[twocolumn,showpacs,preprintnumbers,amsmath,amssymb]{revtex4}

\usepackage{graphicx}
\usepackage{dcolumn}
\usepackage{bm}

\begin{document}
\title{Thermal transport in the hidden-order state of URu$_{2}$Si$_{2}$}
\author{K. Behnia$^{1,2}$, R. Bel$^{1}$, Y. Kasahara$^{2}$, Y. Nakajima$^{2}$,
H. Jin$^{1}$, H. Aubin$^{1}$, K. Izawa$^{2}$, Y. Matsuda$^{2,3}$,
J. Flouquet$^{4}$, Y. Haga$^{5}$, Y. \={O}nuki$^{5}$ and P.
Lejay$^{6}$}
 \affiliation{(1)Laboratoire de Physique Quantique(CNRS), ESPCI, 10 Rue de Vauquelin,
75231 Paris, France \\ (2)Institute for Solid State Physics,
University of Tokyo, Kashiwanoha, Kashiwa, Chiba 277-8581 Japan\\
(3)Department of Physics,
University of Kyoto, Kyoto 608-8502, Japan\\
(4)DRFMC/SPSMS,  Commissariat \`a l'Energie
Atomique, F-38042 Grenoble, France\\
(5)Advanced Science Research Center, Japanese Atomic Energy
Research Institute, Tokai, Ibaraki 319-1195, Japan\\ (6)Centre de
Recherche sur les Tr\`es Basses Temp\'eratures(CNRS), F-38042
Grenoble, France}

\date{March 3, 2005}

\begin{abstract}
We present a study of thermal conductivity in the normal state of
the heavy-fermion superconductor URu$_{2}$Si$_{2}$. Ordering at
18K leads to a steep increase in thermal conductivity and (in
contrast with all other cases of magnetic ordering in
heavy-fermion compounds) to an enhancement of the Lorenz number.
By linking this observation to several other previously reported
features, we conclude that most of the carriers disappear in the
ordered state and this leads to a drastic increase in both
phononic and electronic mean-free-path.
\end{abstract}


\maketitle

Over the years, the phase transition which occurs at T$_{0}\sim$
18 K in URu$_{2}$Si$_{2}$ has become a notorious enigma of
Heavy-Fermion(HF) physics. This phase transition is associated
with a large jump in heat capacity\cite{palstra,schlabitz,maple}
similar to the one observed in several anti-ferromagnetically
ordered HF compounds. On the other hand, and in contrast with the
latter, the magnetic moment in the ordered state appears to be
very weak($\sim 0.03 \mu_{B}$/U)\cite{broholm}. Such a small
magnetic moment is a feature found in many HF compounds. The
puzzle of URu$_{2}$Si$_{2}$ resides in this unique combination.
This is the only case of ordering by heavy electrons with large
anomalies in all macroscopic properties leading to a tiny magnetic
moment.

In order to resolve this apparent paradox, many models have been
proposed\cite{barzykin,santini,sikkema,okuno,ikeda,chandra1,mineev}.
It is widely suspected that there is a hidden order
parameter\cite{shah} distinct from the weak antiferromagnetism.
Several exotic orders have been
imagined\cite{barzykin,santini,ikeda,chandra1}. More recently, a
 feature in the $^{29}$Si NMR data has provided support for
 electronic phase separation in the hidden-order
state\cite{matsuda}. The debate has been mostly focused on the
unusual thermodynamic properties of this ordering. The challenge
for the theory has been to identify the degrees of freedom
corresponding to the huge amount of entropy lost in the
transition. Transport properties have not attracted a comparable
attention. However, as indicated by the recent observation of a
very large Nernst effect in the hidden-order state\cite{bel}, they
may prove to contain interesting information.
\begin{figure}
\resizebox{!}{0.35\textwidth}{\includegraphics{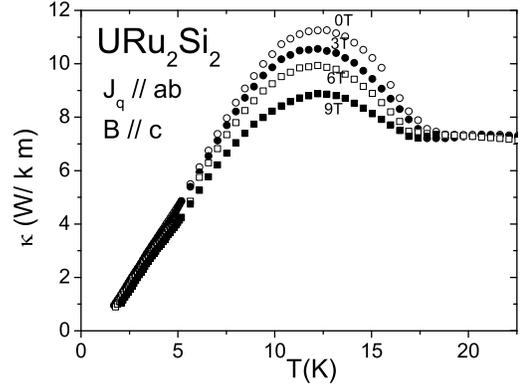}}
\caption{\label{fig1}Temperature dependence of thermal
conductivity of sample 1 for different magnetic fields.}
\end{figure}

In this paper, we report on a study of thermal conductivity in
URu$_{2}$Si$_{2}$ that detects a notable difference between this
compound and all other HF systems which order
anti-ferromagneically. The distinct signature of this phase
transition in thermal transport is a steep increase in the Lorenz
number at the onset of transition. After checking the validity of
the Wiedemann-Franz law in the ordered state, we will argue that
results support a scenario in which most of the electronic
carriers vanish and this leads to an increase in the
mean-free-path of both surviving quasi-particles and heat-carrying
phonons. Thus, the consequences of this phase transition on
thermal transport are strikingly similar to the well-known case of
the superconducting transition in the high-T$_{c}$ cuprates.

This observation highlights the drastic decrease in the carrier
density induced by the hidden order leads in URu$_{2}$Si$_{2}$,
which becomes one order of magnitude lower than in comparable
magnetically-ordered HF compounds. This neglected feature provides
unnoticed constraints for theoretical models.

The two single crystals of URu$_{2}$Si$_{2}$ used in this study
were prepared by Czochralski method in Grenoble and in Tokai. They
were designated as no. 1 (2) with a residual resistivity of
$\rho_{0} \sim$ 10.3(19.5) $\mu\Omega$ cm.
One-heater-two-thermometers set-ups were used to measure both the
longitudinal thermal conductivity (in both samples) and the
transverse thermal conductivity (in sample 2). Cernox chips were
used as thermometers in both set-ups. The thermoelectric (Seebeck
and Nernst) coefficients of sample 1 were also measured using an
identical set-up and recently reported in a separate
communication\cite{bel}.

Fig. 1 displays the thermal conductivity, $\kappa(T)$, of
URu$_{2}$Si$_{2}$ as a function of temperature for different
magnetic fields. The data, measured on sample 1, are similar to
the results obtained for sample 2. As seen in the figure, the
onset of transition at T$_{0}\sim$ 18 K is accompanied with an
enhancement of thermal conductivity leading to the appearance of a
visible maximum of thermal conductivity in the ordered state. As
seen in the figure, this upturn in $\kappa(T)$ is reduced by the
application of a magnetic field. Since the magnetic field is known
both to gradually destroy the ordered moment and to reduce
T$_{0}$\cite{mentink,bourdarot}, the latter observation is not
surprising.

In this regard, the case of URu$_{2}$Si$_{2}$ appears identical to
other compounds studied in the vicinity of a magnetic order. This
is the case of UPd$_{2}$Al$_{3}$ (which orders
antiferromagnetically at T$_{N}\sim$ 14 K)\cite{hiroi},
CeRhIn$_{5}$ (T$_{N}$=3.8K)\cite{paglione} as well as rare-earth
compounds of the generic formula RB$_{6}$ (with R=Pr, Nd, Gd and 4
K $<$ T$_{N} <$ 16 K)\cite{sera}. In all these cases, heat
transport in the ordered state improves due to the sudden freezing
of a major scattering mechanism of the heat carriers.

\begin{figure}
\resizebox{!}{0.35\textwidth}{\includegraphics{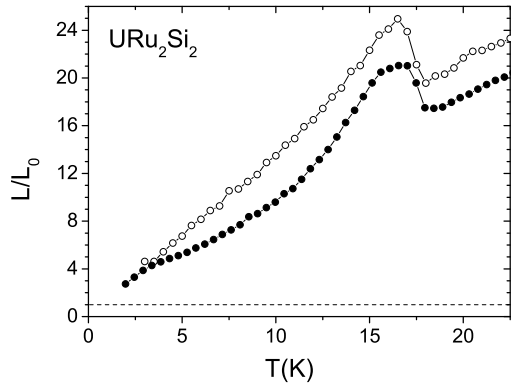}}
\caption{\label{fig2} The zero-field Lorenz number normalized to
the Sommerfeld value as a function temperature for sample 1 (solid
circles) and sample 2 (empty circles).}
\end{figure}

However, URu$_{2}$Si$_{2}$ presents a unique feature which becomes
visible by comparing the conduction of heat, $\kappa$, and charge,
$\sigma$, and contrasting the change in each induced by the onset
of ordering. One convenient method for such a comparison is to
focus on the temperature dependence of the Lorenz number, a ratio
of thermal to charge conductivities: L=$\frac{\kappa}{\sigma T}$.
According to the Wiedemann-Franz(WF) law, in absence of lattice
conductivity and inelastic scattering of electrons, this number
becomes equal to the Sommerfeld value,
L$_{0}$=$\frac{\pi^{2}}{3}(\frac{k_{B}}{e})^2=24.4 nW/(K^{2} m)$.
Fig. 2 displays the temperature dependence of the normalized
Lorenz number, $\frac{L}{L_{0}}$ in the two samples of
URu$_{2}$Si$_{2}$ used in this study. As seen in the figure, in
both cases ordering leads to a sudden \emph{increase} in
$\frac{L}{L_{0}}$. In other words, thermal conduction, even after
normalization to the charge transport, still displays an
enhancement. This is in sharp contrast with the other compounds
mentioned above. In those cases, ordering leads to a decrease in
$\frac{L}{L_{0}}$: the enhancement in thermal conduction is not
large enough to match the increase in the charge transport
canal\cite{hiroi,sera,paglione}.

In order to explore the possible origin of this singular behavior
of URu$_{2}$Si$_{2}$, let us begin by separating the effect of the
phase transition on different types of heat carriers. As seen in
Fig. 2, the large magnitude of $\frac{L}{L_{0}}$ ($\sim$ 18) at
the onset of transition indicates that the contribution of the
quasi-particles to heat transport constitutes a tiny fraction of
the total thermal conductivity. The situation is similar in
UPd$_{2}$Al$_{3}$ where $\frac{L(T=T_{N})}{L_{0}}$ $\sim$
11\cite{hiroi}. However, this is not the case of PrB$_{6}$,
NdB$_{6}$, GdB$_{6}$ or CeRhIn$_{5}$. In the latter systems, with
quasi-particles carrying most or all of heat, the observed
decrease in $\frac{L}{L_{0}}$ is undoubtedly due to a change in
the inelastic scattering of electrons. Above T$_{N}$, spin
fluctuations scatter conduction electrons and their sudden
freezing by the onset of ordering leads to a steep increase in
conductivity\cite{sera,paglione}. Now, inelastic scattering is
 more efficient in impeding the transport of heat than
charge, since those scattering events which imply little change in
the momentum of the scattered quasi-particle leave a much stronger
signature in thermal resistance. In this context, a sudden drop in
$\frac{L}{L_{0}}$ with ordering is a signature of more frequent
small wave-vector scattering events in the ordered state. In other
words, the presence of magnetic fluctuations above T$_{N}$ tends
to amplify the relative weight of large-\textbf{q} scattering and
to rectify the excess in thermal resistivity produced by inelastic
e-e scattering. Interestingly, this picture seems relevant even
for UPd$_{2}$Al$_{3}$. In spite of the much smaller relative
weight of quasi-particles in heat transport (which account for
less than ten percent of the total), the transition is accompanied
with a \emph{reduction} of $\frac{L}{L_{0}}$\cite{hiroi}.
Therefore, one is brought to explore the possible reasons which
makes the case of URu$_{2}$Si$_{2}$ so different. Why does the
occurrence of the hidden order lead to an excessive enhancement of
thermal conductivity?

One hypothetic possibility is the existence of an exotic heat
transport introduced by the hidden order. In order to check this,
we have measured the Righi-Leduc (or the thermal Hall) effect in
the ordered state of URu$_{2}$Si$_{2}$. This effect, which refers
to the emergence of a transverse thermal gradient in response to a
longitudinal heat current (and in presence of a perpendicular
magnetic field) is associated with a finite value of the
off-diagonal thermal conductivity tensor $\kappa_{xy}$. It has
been employed successfully to separate the electronic and lattice
components of heat conduction in the superconducting state of
YBa$_{2}$Cu$_{3}$O$_{7-\delta}$\cite{krishana}. Among heat
carriers, only those which are skew-scattered in presence of a
magnetic field are expected to contribute to  $\kappa_{xy}$. A
verification of the Wiedemann-Franz correlation between
$\kappa_{xy}$ and $\sigma_{xy}$ has been reported for
copper\cite{zhang}.
\begin{figure}
\resizebox{!}{0.35\textwidth}{\includegraphics{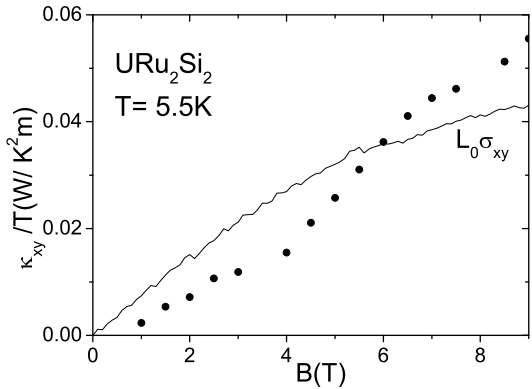}}
\caption{\label{fig3}The field dependence of
$\frac{\kappa_{xy}}{T}$ at T=5.5 K in sample 2. The solid line
represents the field-dependence of $\sigma_{xy}$ in the same
sample multiplied by the Sommerfeld value.}
\end{figure}

As seen in Fig.3, at T=5.5K, that is well below T$_{0}$, the
magnitude of $\kappa_{xy}/T$ is comparable with
L$_{0}\sigma_{xy}$. In other words, in the field range extending
from 0 to 9T, the ratio $(\kappa_{xy}/(\sigma_{xy} T L_{0})$
remains between 0.6 and 1.2.  The apparent non-linear
field-dependence of $\kappa_{xy}$ is presumably due to the
temperature instability during field sweeps which can lead to an
uncertainty of 30 percent on $\Delta_{y}T$ (which is of the order
of a few mK). Moreover, the magnetoresistance of the regulating
Cernox thermometer was not corrected. Using the technical
literature data\cite{heine}, we estimate that at 5K and 9T, it can
lead to an overestimation of $\kappa_{xy}$ by 4 percent. If the
hidden-order state was host to any unconventional type of heat
carriers exposed to skew scattering, then one would expect a
$(\kappa_{xy}/(\sigma_{xy} T )$ significantly larger than $L_{0}$.
Even with the level of experimental uncertainty achieved here, it
appears safe to conclude that this is not the case. If there is
any heat transport by magnetic excitations, it cannot be
distinguished from lattice heat transport, at least at this stage.
Therefore, in the following discussion, any contribution to heat
transport by bosonic excitations would be addressed as part of the
conventional phonon heat conductivity.

If heat conduction in URu$_{2}$Si$_{2}$ is the sum of electronic
($\kappa_{e}$) and lattice ($\kappa_{ph}$) components as usual,
then let us separate them in order to see what sets the phase
transition occurring at 18K apart. Assuming $\kappa_{e} = L \sigma
T$, i.e. supposing the validity of the WF law for the electronic
contribution to thermal conductivity, one obtains $\kappa_{e}(T)$
and  $\kappa_{ph}(T)$ as displayed in Fig. 4. The striking feature
of the figure is the enhancement of $\kappa_{ph}$ below T$_{0}$.
Note that If the $\frac{L}{L_{0}}$ drops for the electronic
component, as observed in other systems, then the extracted
enhancement in the lattice contribution would become even
stronger. We will argue below that this feature, exclusive to
URu$_{2}$Si$_{2}$, is a consequence of the vanishing of most of
the itinerant electrons and a concomitant decrease in the
electronic scattering of phonons.

The opening of an energy gap upon ordering in URu$_{2}$Si$_{2}$
was detected as early as the discovery of this phase transition.
An activated behavior is clearly resolved in the temperature
dependence of both resistivity and the specific
heat\cite{palstra,schlabitz,maple}. The magnitude of the energy
gap extracted from these measurements (50-110K) is somewhat larger
than the gap observed in spin excitation spectrum (1.8 meV $\sim$
21 K)\cite{broholm,bourdarot}. The quantification of the fraction
of the Fermi surface destroyed by the opening of this gap,
however, is less straightforward. By monitoring the change in the
magnitude of the linear electronic specific heat ($\gamma
=C_{el}/T$), Fisher and co-workers estimated that the fraction of
the Fermi surface removed is 31 percent\cite{fisher}, not very
different from earlier estimations employing the same
method\cite{maple}. Now, the change in $\gamma$ induced by AF
ordering in UPd$_{2}$Al$_{3}$ (from 210 mJ/K$^{2}$ above T$_{N}$
to 150 below) implies the removal of a comparable fraction of the
Fermi Surface\cite{geibel}. However, the consequences of ordering
for thermal conductivity in the two systems are visibly different.

If the transition affects both the effective mass and the density
of carriers, then the change in specific heat does not simply
reflect the fraction of Fermi Surface lost. There are two distinct
experimental observations indicating that that the change in
$\gamma$ underestimates the fraction of the Fermi Surface lost in
the transition in URu$_{2}$Si$_{2}$. i) The five-fold jump in the
Hall coefficient R$_{H}$ induced by the
transition\cite{schoenes,bel}, which (taken at its face value)
reflects a large decrease in carrier density; ii) The three-fold
increase in the linear term of the thermopower, $S/T$, which
points to an enhancement of \emph{the entropy per carrier} in the
ordered state\cite{bel}. In such a case, the change in entropy per
volume monitored by $\gamma$ is much smaller than the change in
carrier density. Neither of these occur in the case of
UPd$_{2}$Al$_{3}$.
\begin{figure}
\resizebox{!}{0.35\textwidth}{\includegraphics{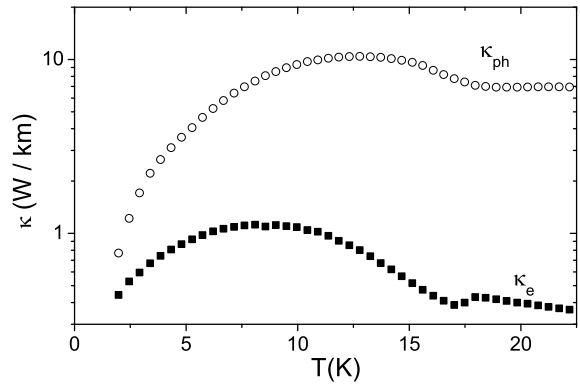}}
\caption{\label{fig4}  Lattice ($\kappa_{ph}$) and
electronic($\kappa_{e}$) components of thermal conductivity in
sample 1, assuming $\kappa_{e} = L_{0} \sigma T$ in the whole
temperature range.}
\end{figure}

Moreover, by comparing the physical properties of
URu$_{2}$Si$_{2}$ and UPd$_{2}$Al$_{3}$ at low temperatures, one
finds three independent lines of evidence suggesting that the
carrier density in the former is one order of magnitude smaller
than the latter. a) The Hall coefficient in the zero-temperature
limit, remarkably large in URu$_{2}$Si$_{2}$ (R$_{H} \sim
10^{-8}m^{3}/C$ corresponding to 0.05 carriers per U in a simple
one-band picture\cite{schoenes,bel}), exceeds by a factor of
twenty the same quantity in UPd$_{2}$Al$_{3}$\cite{huth}. Note
that neither multi-band effects (which would eventually reduce the
total R$_{H}$) or skew scattering (estimated in the
zero-temperature limit using the Pauli
susceptibility\cite{schoenes}) can explain the magnitude of
R$_{H}$ in URu$_{2}$Si$_{2}$. b) In dHvA studies, a Dingle
temperature of similar magnitude (T$_{D} \sim$ 0.2K) was obtained,
in spite of the fact that the residual resistivity of the
URu$_{2}$Si$_{2}$ sample (with $\rho_{0} \sim 9.5 \mu \Omega$ cm)
studied was much higher than the UPd$_{2}$Al$_{3}$ one ($\rho_{0}
\sim 1.4 \mu \Omega$ cm). In other words, the same carrier
mean-free-path corresponds to an electric conductivity which is
almost one order of magnitude lower in
URu$_{2}$Si$_{2}$\cite{dhva}. c) The superconducting penetration
depth, $\lambda$ is almost 2.5 times larger in URu$_{2}$Si$_{2}$
than in UPd$_{2}$Al$_{3}$\cite{amato}. Since
$\frac{1}{\lambda^{2}}\propto\frac{n_{s}}{m^{*}}$, this implies
that the ratio of the superfluid density, n$_{s}$, to the
effective mass, m$^{*}$, is more than six times larger in the
former compound.

If ordering in URu$_{2}$Si$_{2}$ leads to the removal of
nine-tenth of the Fermi surface as suggested by the
above-mentioned data (and imaginable in a Spin Density Wave
scenario), its intriguing signature on thermal transport will find
a natural explanation. Lattice thermal conductivity is known to
increase abruptly in many Charge Density Wave transitions because
of the vanishing of electronic scatterers\cite{nunez,lopes}.
Sizable increase in phonon thermal conductivity due to the opening
of a superconducting gap\cite{belin} is not unusual either. The
case of URu$_{2}$Si$_{2}$ is more intriguing as it leads to an
increase in \emph{both }thermal and electric conductivities in
spite of the loss of a huge fraction of charged carriers. In other
words, the partial destruction of the Fermi surface leads to an
increase in the scattering time of both phonons and
electrons\cite{scattering}. Thermal transport in this context
presents a curious similarity with the more familiar case of
cuprates. In YBa$_{2}$Cu$_{3}$O$_{7-\delta}$, the opening of the
d-wave superconducting gap leads to an enhancement of both
phononic and electronic components of thermal
conductivity\cite{krishana}.

The diluted carrier concentration in URu$_{2}$Si$_{2}$ may prove
to be an important piece of the puzzle. Until now, the debate has
focused on  the small magnetic moment of 0.03 $\mu_{B}$/U without
considering the density of itinerant electrons per uranium which
is also unusually small. An intimate connection between these two
properties remains an open question. They may be two distinct
consequences of the Fermi surface nesting at T$_{0}$. Further
exploration of transport properties under pressure, where the
hidden order is replaced by a large moment AF state is clearly
desirable. Theoretically, thermal transport by nodal
quasi-particles in an unconventional density wave
state\cite{ikeda} appears as an interesting subject to explore.

In summary, our study of heat transport in URu$_{2}$Si$_{2}$
detected a drastic enhancement in lattice thermal conductivity
consequent to the loss of a large fraction of the Fermi surface.
Both the electronic and phonon lifetime are enhanced in the
ordered state which appears to be associated with a remarkably low
level of carrier concentration.

K. B. acknowledges the hospitality of the University of Tokyo,
where this research was partially carried out.


\begin{thebibliography}{99}
\bibitem{scattering} The increase in the electronic mean-free-path
induced by the transition is also reflected in the 30-fold
enhancement of the Hall angle below T$_{0}$\cite{bel}.
\end{thebibliography}
\end{document}